\documentclass[sigconf,nonacm]{acmart}

\usepackage{graphicx}
\usepackage{amsmath,amssymb,amsfonts}   
\usepackage{glossaries}                 
\usepackage{enumitem}                   
\usepackage{makecell}                   
\usepackage{pifont}                     
\usepackage{cleveref}                   
\usepackage{color-edits}                
\usepackage{multirow}                   
\usepackage[table]{xcolor}              
\usepackage{float}                      
\usepackage{caption}                    
\usepackage{booktabs}                   
\usepackage{balance}					

\definecolor{lightgray}{HTML}{CCCCCC}

\newif\ifnb     
\nbfalse
\nbtrue

\newif\ifcom    
\comtrue

\newif\ifarx    
\arxfalse

\crefname{algocf}{Alg.}{Algs.}
\crefname{figure}{Fig.}{Figs.}
\crefname{equation}{Eq.}{Eqs.}

\addauthor[Patrick]{pi}{teal}
\renewcommand{\pi}[1]{\ifcom\picomment{#1}\fi}

\newcommand{\il}{\raisebox{-0.125em}{\includegraphics[height=0.9em]{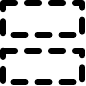}}}
\newcommand{\loi}{\raisebox{-0.125em}{\includegraphics[height=0.9em]{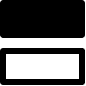}}}
\newcommand{\lol}{\raisebox{-0.125em}{\includegraphics[height=0.9em]{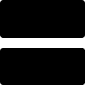}}}

\newcommand{\wdi}{\raisebox{-0.125em}{\includegraphics[height=0.9em]{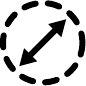}}}
\newcommand{\wds}{\raisebox{-0.125em}{\includegraphics[height=0.9em]{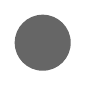}}}
\newcommand{\wdl}{\raisebox{-0.125em}{\includegraphics[height=0.9em]{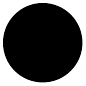}}}

\newcommand{\wu}{\raisebox{-0.125em}{\includegraphics[height=0.9em]{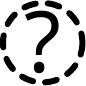}}}
\newcommand{\wur}{\raisebox{-0.125em}{\includegraphics[height=0.9em]{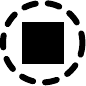}}}
\newcommand{\wum}{\raisebox{-0.125em}{\includegraphics[height=0.9em]{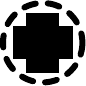}}}

\newcommand{\sfi}{\raisebox{-0.125em}{\includegraphics[height=0.9em]{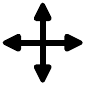}}}
\newcommand{\sfr}{\raisebox{-0.125em}{\includegraphics[height=0.9em]{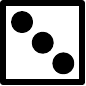}}}
\newcommand{\sfa}{\raisebox{-0.125em}{\includegraphics[height=0.9em]{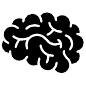}}}

\newcommand{\github}{\ifnb https://github.com/spcl/nw-design-for-wsi \else https://github.com/anon8274/wsi (anonymized for blind review)\fi}

\newacronym{f2f}{F2F}{face-to-face}
\newacronym{f2b}{F2B}{face-to-back}
\newacronym{d2d}{D2D}{die-to-die}
\newacronym{tsv}{TSV}{through-silicon via}
\newacronym{phy}{PHY}{physical layer}
\newacronym{wow}{WoW}{wafer-on-wafer}
\newacronym{kgd}{KGD}{known-good die}
\newacronym{hb}{HB}{hybrid bond}
\newacronym{ml}{ML}{machine learning}
\newacronym{hpc}{HPC}{high-performance computing}
\newacronym{gpc}{GPC}{graphics processing cluster}
\newacronym{loi}{LoI}{logic-on-interconnect}
\newacronym{lol}{LoL}{logic-on-logic}
\newacronym{scb}{SCB}{simple cycle-breaking}
\newacronym{noc}{NoC}{network-on-chip}
\newacronym{ici}{ICI}{inter-chip interconnect}
\newacronym{llm}{LLM}{large language model}
\newacronym{wsi}{WSI}{wafer-scale integration}
\newacronym{moe}{MoE}{mixture-of-experts}

\setcopyright{acmlicensed}
\copyrightyear{2025}
\acmYear{2025}
\acmDOI{XXXXXXX.XXXXXXX}
\acmConference[DAC '26]{Design Automation Conference}{July 26--29, 2025}{Long Beach, CA}
\acmISBN{978-1-4503-XXXX-X/2018/06}

\begin{document}

\title{Network Design for Wafer-Scale Systems with Wafer-on-Wafer Hybrid Bonding}

\ifnb
	\author{Patrick Iff}
	\affiliation{%
	  \institution{ETH Zurich}
	  \city{Zurich}
	  \country{Switzerland}}
	\email{iffp@inf.ethz.ch}

	\author{Tommaso Bonato}
	\affiliation{%
	  \institution{ETH Zurich}
	  \city{Zurich}
	  \country{Switzerland}}

	\author{Maciej Besta}
	\affiliation{%
	  \institution{ETH Zurich}
	  \city{Zurich}
	  \country{Switzerland}}

	\author{Luca Benini}
	\affiliation{%
	  \institution{ETH Zurich}
	  \city{Zurich}
	  \country{Switzerland}}

	\author{Torsten Hoefler}
	\affiliation{%
	  \institution{ETH Zurich}
	  \city{Zurich}
	  \country{Switzerland}}
	\email{htor@inf.ethz.ch}
\else
\fi

\ifnb
\renewcommand{\shortauthors}{Iff et al.}
\else
\renewcommand{\shortauthors}{Anonymous et al.}
\fi

\begin{abstract}
	Transformer-based large language models are increasingly constrained by data movement as communication bandwidth drops sharply beyond the chip boundary.
Wafer-scale integration using wafer-on-wafer hybrid bonding alleviates this limitation by providing ultra-high bandwidth between reticles on bonded wafers.
In this paper, we investigate how the physical placement of reticles on wafers influences the achievable network topology and the resulting communication performance.
Starting from a 2D mesh-like baseline, we propose four reticle placements (\emph{Aligned}, \emph{Interleaved}, \emph{Rotated}, and \emph{Contoured}) that improve throughput by up to 250\%, reduce latency by up to 36\%, and decrease energy per transmitted byte by up to 38\%.
\\ 

	\vspace{-1em}
\end{abstract}

\ifnb
\fi

\ifnb
\fi

\maketitle

\vspace{-0.5em}
\section{Introduction}
\label{sec:intro}

Transformer-based \glspl{llm} power today’s most advanced AI systems, enabling breakthroughs in reasoning, generation, and multimodal understanding.
Training these models is increasingly constrained by data movement \cite{datamovement}.
Communication bandwidth declines sharply across hierarchy levels, from on-chip interconnects (multiple TB/s) to intra-node connections such as NVLink ($\approx$ 900 GB/s) and inter-node fabrics like NVIDIA’s NDR InfiniBand ($\approx$ 100 GB/s) \cite{wafer-scale-survey}.
For decades, transistor scaling continually increased on-chip compute density and mitigated communication bottlenecks, but the slowdown of Moore’s law and the end of Dennard scaling has largely curtailed these gains.
\Gls{wsi} provides an alternative path by scaling up the physical chip size itself, enabling entire wafers to function as unified substrates with high-bandwidth internal communication.

Wafer-on-wafer integration with hybrid bonding \cite{hybrid-bonding} is a promising and commercially available approach for building wafer-scale systems, exemplified by TSMC’s SoIC-WoW \cite{tsmc-soic-wow}.
In these systems, two silicon wafers are bonded \gls{f2f} using hybrid bonding, enabling high-density inter-wafer connections.
Unlike reticle stitching, adjacent reticles on the same wafer cannot communicate directly; instead, reticles must be placed so that connecting any two overlapping reticles on opposite wafers yields a fully connected system.
This integration scheme introduces a new and unexplored design space for on-chip interconnects, where the physical placement of reticles on wafers dictates the achievable network topologies, a key factor in communication performance.

In this paper, we explore how to place reticles on the top and bottom wafers to achieve efficient network topologies.
Starting from a 2D mesh-like topology with up to four neighbors per reticle, we investigate how alternative reticle placements reduce the average path length by enabling up to seven neighbors per reticle.
We introduce four novel reticle placements (\emph{Aligned}, \emph{Interleaved}, \emph{Rotated}, and \emph{Contoured}) that are tailored to different architectural configurations and present different trade-offs in terms of design and manufacturing complexity, and performance.
Our evaluation shows that these placements improve overall throughput by up to 250\%, while reducing average packet latency and energy per transmitted byte by up to 36\% and 38\%, respectively, compared to the baseline 2D mesh-like topology.

\section{Background on Wafer-Scale Integration}
\label{sec:back}

\subsection{Approaches to Wafer-Scale Integration}
\label{ssec:back-approaches}

Several approaches exist to achieve wafer-scale integration.
Tesla's Dojo \cite{tesla-dojo}, integrates $25$ silicon dies of $645$ mm$^2$ into a single wafer-scale system by placing individual chiplets on a fan-out wafer, a method known as \textbf{chiplet-based wafer-scale integration}.
Cerebras \cite{cerebras} employs \textbf{field stitching} \cite{field-stitching} to overcome the reticle limit of $26\times33$ mm.
Field stitching introduces a small, intentional overlap between neighboring reticle exposures to align circuit patterns across seams and create continuous wires that cross reticle boundaries.
In this work, we focus on a third approach, \textbf{wafer-on-wafer hybrid bonding}, as offered by TSMC's SoIC-WoW process \cite{tsmc-soic-wow}, where two wafers are patterned with reticles and bonded \gls{f2f} to form a single wafer-scale chip.
Reticles on the same wafer cannot be connected directly, but by interleaving reticles on both wafers and vertically connecting them through \glspl{hb}, a fully connected network among all reticles is built.

\subsection{Wafer-on-Wafer Hybrid Bonding}

A key advantage of wafer-on-wafer hybrid bonding over chiplet-based \gls{wsi} is the extremely small pitch of \glspl{hb}, which is below $10\,\mu m$ in production \cite{chiplet-interconnects} and reaches $1\,\mu m$ in research prototypes \cite{stacking-3-wafers}.
Unlike \gls{d2d} links in chiplet-based wafer-scale integration, which require dedicated, area- and power-intensive \glspl{phy} on both ends for protocol, frequency, and voltage translation, hybrid bonding requires no \glspl{phy} because its electrical characteristics resemble those of the upper metal layers.
Moreover, while \gls{d2d} link bandwidth is typically limited by the number of available microbumps \cite{hexamesh, foldedhexatorus}, the fine pitch of hybrid bonding shifts the bottleneck to wire routing from the \glspl{hb} to the router or to the router area itself.
We use the term \textit{vertical connector} to describe a collection of \glspl{hb} that enable one link between wafers.
To build a link between two reticles, the reticles must be on opposite wafers, and their vertical connectors must be precisely aligned.

\section{Architecture Overview}
\label{sec:arch}

In this section, we describe the different architectural choices and routing strategies we consider in this work.

\subsection{System Architecture}
\label{ssec:arch-system}

We target \gls{ml} and \gls{hpc} workloads suitable for acceleration by GPU-like architectures.
Each $26 \times 33$~mm reticle contains a GPU with eight \glspl{gpc} and local SRAM.

\underline{\il~Integration Level:}
We explore two levels of vertical integration.
In \textbf{\loi~\gls{loi}}, only the top wafer contains compute reticles (i.e., GPUs), while the bottom wafer serves purely as an interconnect layer.
Limiting the system to a single compute wafer directly attached to the heat sink simplifies thermal management and power delivery.
The second integration level, called \textbf{\lol~\gls{lol}}, places compute reticles on both wafers, with the interconnect integrated into the compute reticles.
While technically feasible today, power and thermal constraints remain major challenges.
We expect \gls{lol} to become viable within a few years through improved power efficiency or advanced cooling, such as thermal through-silicon vias \cite{thermal-tsv} or microfluidic cooling \cite{microfluidic-cooling}.

\underline{\wdi~Wafer Diameter:}
We analyze \textbf{\wdl~300~mm} wafers, which represent the current mainstream in semiconductor manufacturing, and \textbf{\wds~200~mm} wafers, still used in some older fabs.

\underline{\wu~Wafer Utilization:}
We analyze two levels of wafer utilization.
The common approach in literature is to assemble a \textbf{\wur~rectangular} 2D grid of chiplets or reticles on a wafer \cite{tesla-dojo, wsc-llm, data-center-in-cabinet}.
We also consider the case where wafer utilization is \textbf{\wum~maximized} by tightly packing the largest possible number of reticles onto the wafer.
This enables more efficient use of silicon and, importantly, a system with increased compute capabilities and higher integration density.
Since wafer-on-wafer hybrid bonding removes the need for dicing streets between reticles, we omit inter-reticle spacing from our model, as the remaining micrometer-scale spacing does not noticeably affect the results.

\subsection{Network Architecture}

We assume a packet-switched network with wormhole routing and credit-based flow control.
Following Yin et al. \cite{abstract-router-model}, who abstract the global network into a single router to optimize a chiplet's local network in isolation, we abstract each compute reticle's local network into a single router to optimize the wafer-scale network independently.
Thus, a compute reticle is modeled as one router connecting all \glspl{gpc} and the reticle’s vertical connectors.
For interconnect reticles in \loi~\gls{loi}, we explicitly model routers and their connecting links.

The routing algorithm consists of two components: the routing function and the selection function, both invoked when a packet traverses a router.
The routing function returns a list of output ports that ensure deadlock- and livelock-freedom, from which the selection function chooses one.
Our routing algorithm applies Dijkstra's algorithm \cite{dijkstra} to return only shortest paths, ensuring progress toward the destination and thus guaranteeing livelock-freedom.
It also employs the \gls{scb} algorithm \cite{scb-algorithm}, a turn model \cite{turn-model} variant for arbitrary topologies, to guarantee deadlock-freedom.

\underline{\sfi~Selection Function:}
We evaluate two different selection functions: \textbf{\sfr~random}, which chooses an output port at random, and local \textbf{\sfa~adaptive}, which selects the port towards the router with the most available space in its input buffer.
Since the input buffer occupancy of adjacent routers is available via credit-based flow control, the adaptive selection function can be implemented without additional overhead to communicate congestion information.

The network topology is arguably the most critical aspect of the network architecture. In wafer-on-wafer hybrid bonding, links can connect only overlapping reticles on opposite wafers; thus, the topology is dictated by the reticle placement, whose optimization is the central contribution of this work.

\vspace{-0.75em}
\section{Optimization of Reticle Placement}
\label{sec:opt}

\begin{figure*}[h]
\centering
\captionsetup{justification=centering}
\includegraphics[width=1.0\textwidth]{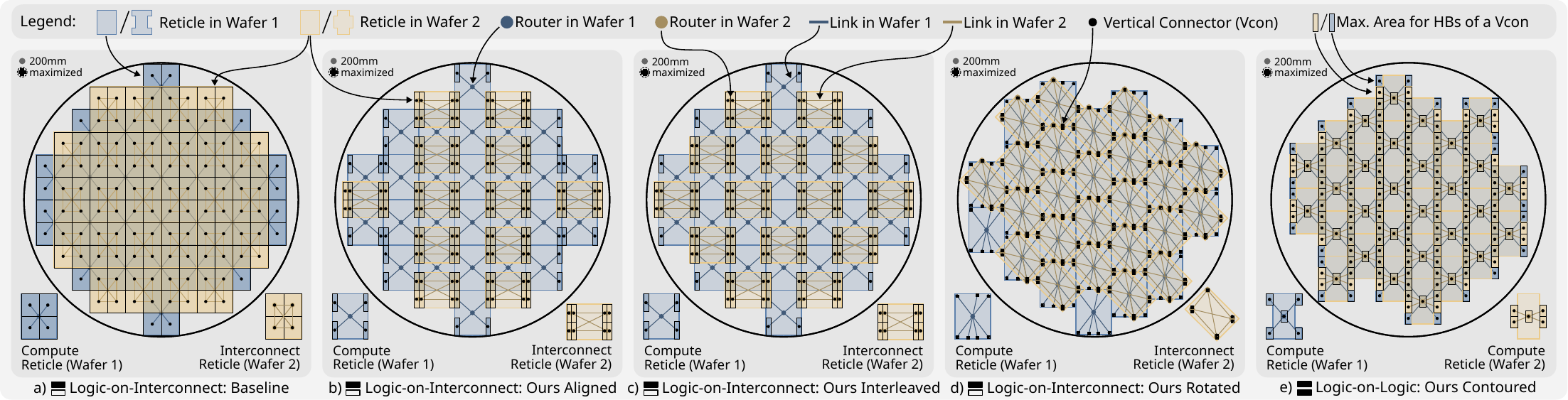}
\vspace{-2.25em}
\caption{(\textsection \ref{sec:opt}) Different methods of building wafer-scale systems by optimizing the placement of reticles on the wafers.}
\vspace{-1.0em}
\label{fig:system-visualizations}
\end{figure*}

In this section, we optimize the network topology.
Previous work on network topologies, from supercomputers \cite{dragonfly, slimfly} to \glspl{noc} \cite{slimnoc, ruche-networks} and \glspl{ici} \cite{hexamesh, foldedhexatorus, butterdonut, kite}, focused on minimizing network diameter and average path length.
Reducing these metrics decreases latency, mitigates congestion, and increases throughput.
In wafer-on-wafer hybrid bonding systems, only links between overlapping reticles can be implemented, so optimizing the topology requires optimizing reticle placement.

\begin{table}[h]
\vspace{-1.0em}
\centering
\footnotesize
\setlength{\tabcolsep}{2.5pt}
\caption{(\textsection \ref{sec:opt}) Comparison of different reticle placements.}
\vspace{-1.5em}
\label{tab:overview}
\begin{tabular}{lllcccccccc}
\toprule
\rotatebox{90}{\il~Integration Level} & \rotatebox{90}{\wdi~Wafer Diameter} & \rotatebox{90}{\wu~Wafer Utilization} & Placement & \rotatebox{90}{\makecell[l]{Number of Compute\\Reticles (26$\times$33mm)}} & \rotatebox{90}{\makecell[l]{Number of Interconnect\\Reticles ($\approx$26$\times$33mm)}} & \rotatebox{90}{\makecell[l]{Radix of\\Compute Reticles}} & \rotatebox{90}{\makecell[l]{Radix of\\Interconnect Reticles}} & \rotatebox{90}{\makecell[l]{Network\\Diameter}} & \rotatebox{90}{\makecell[l]{Average Path\\Length (Hops)}} & \rotatebox{90}{\makecell[l]{Total Bisection\\Bandwidth}} \\
\midrule
\multirow{16}{*}{\rotatebox{90}{\makecell{\loi~Logic on Interconnect}}} & \multirow{8}{*}{\rotatebox{90}{\makecell{\wds~200mm}}} & \multirow{4}{*}{\rotatebox{90}{\makecell{\wur~Rec.\mbox{\hspace{0.2em}}}}} & Baseline & 20 & 26 & 4 & 4 & 8 & 4.08 & 16.00 \\
 &  &  & Ours Aligned & 20 & 10 & 4 & 6 & 6 & 3.30 & 16.00 \\
 &  &  & Ours Interleaved & 20 & 12 & 4 & 6 & 8 & 3.44 & 16.00 \\
 &  &  & Ours Rotated & 20 & 20 & 7 & 7 & 6 & 2.84 & 32.00 \\
\cline{3-11}
 &  & \multirow{4}{*}{\rotatebox{90}{\makecell{\wum~Max.\mbox{\hspace{0.2em}}}}} & Baseline & 26 & 26 & 4 & 4 & 12 & 4.80 & 16.00 \\
 &  &  & Ours Aligned & 26 & 12 & 4 & 6 & 10 & 3.91 & 16.40 \\
 &  &  & Ours Interleaved & 26 & 14 & 4 & 6 & 10 & 3.89 & 16.00 \\
 &  &  & Ours Rotated & 27 & 25 & 7 & 7 & 6 & 3.20 & 38.00 \\
\cline{2-11}
 & \multirow{8}{*}{\rotatebox{90}{\makecell{\wdl~300mm}}} & \multirow{4}{*}{\rotatebox{90}{\makecell{\wur~Rec.\mbox{\hspace{0.2em}}}}} & Baseline & 49 & 56 & 4 & 4 & 12 & 6.44 & 27.20 \\
 &  &  & Ours Aligned & 49 & 28 & 4 & 6 & 12 & 5.53 & 28.00 \\
 &  &  & Ours Interleaved & 49 & 26 & 4 & 6 & 12 & 5.57 & 24.00 \\
 &  &  & Ours Rotated & 48 & 48 & 7 & 7 & 10 & 4.19 & 47.60 \\
\cline{3-11}
 &  & \multirow{4}{*}{\rotatebox{90}{\makecell{\wum~Max.\mbox{\hspace{0.2em}}}}} & Baseline & 64 & 63 & 4 & 4 & 18 & 7.45 & 26.00 \\
 &  &  & Ours Aligned & 64 & 31 & 4 & 6 & 14 & 5.83 & 31.20 \\
 &  &  & Ours Interleaved & 64 & 31 & 4 & 6 & 14 & 6.04 & 28.20 \\
 &  &  & Ours Rotated & 66 & 63 & 7 & 7 & 10 & 4.76 & 64.20 \\
\cline{1-11}
\multirow{8}{*}{\rotatebox{90}{\makecell{\lol~Logic on Logic}}} & \multirow{4}{*}{\rotatebox{90}{\makecell{\wds200mm\mbox{\hspace{0.4em}}}}} & \multirow{2}{*}{\rotatebox{90}{\makecell{Rec.\mbox{\hspace{0.2em}}}}} & Baseline & 46 & 0 & 4 & - & 10 & 4.40 & 16.00 \\
 &  &  & Ours Contoured & 40 & 0 & 5 & - & 8 & 3.52 & 16.00 \\
\cline{3-11}
 &  & \multirow{2}{*}{\rotatebox{90}{\makecell{Max.\mbox{\hspace{0.2em}}}}} & Baseline & 52 & 0 & 4 & - & 12 & 4.71 & 16.00 \\
 &  &  & Ours Contoured & 54 & 0 & 5 & - & 10 & 3.93 & 21.20 \\
\cline{2-11}
 & \multirow{4}{*}{\rotatebox{90}{\makecell{\wdl300mm\mbox{\hspace{0.4em}}}}} & \multirow{2}{*}{\rotatebox{90}{\makecell{Rec.\mbox{\hspace{0.2em}}}}} & Baseline & 105 & 0 & 4 & - & 14 & 6.66 & 27.20 \\
 &  &  & Ours Contoured & 96 & 0 & 5 & - & 12 & 5.20 & 28.00 \\
\cline{3-11}
 &  & \multirow{2}{*}{\rotatebox{90}{\makecell{Max.\mbox{\hspace{0.2em}}}}} & Baseline & 127 & 0 & 4 & - & 20 & 7.42 & 25.60 \\
 &  &  & Ours Contoured & 132 & 0 & 5 & - & 16 & 6.01 & 36.00 \\
\bottomrule
\end{tabular}
\vspace{-1.5em}
\end{table}

Our approach is to minimize the average path length by maximizing the number of overlapping reticles between the two wafers (i.e., the network radix).
While we present results for reticles at the lithographic limit of  $26 \times 33$~mm, all optimization techniques apply to other reticle sizes as well.
\Cref{tab:overview} lists the reticle count, radix, diameter, average path length, and estimated bisection bandwidth (averaged over ten METIS \cite{metis} runs with different random seeds) for all proposed reticle placements.
Diameter and average path length use reticle-to-reticle rather than router-to-router hops.

\subsection{Placements for \loi~Logic-on-Interconnect}
\label{ssec:opt-loi}

\paragraph{Constraints} 
We assume that to minimize design cost and maximize manufacturing efficiency, all reticles on a given wafer are required to be identical.

\paragraph{Baseline} 
Most prior work on wafer-scale networks \cite{wsc-llm,fred,network-on-wafer,data-center-in-cabinet,waferscale-gpu,ucla-uiuc} focuses on chiplet-based wafer-scale integration rather than wafer-to-wafer hybrid bonding, leaving no established baseline for this unexplored network design space.
Since many chiplet-based systems use a 2D mesh topology as a baseline, we adopt a system that approximates a 2D mesh.
In this baseline, the reticles on the interconnect wafer are shifted by half a reticle width and height so that each interconnect reticle connects to four neighboring compute reticles, and vice versa (see \Cref{fig:system-visualizations}a).
Each interconnect reticle contains a fully connected radix-$4$ network topology.
Note that the resulting topology\footnote{Visualizations of all network topologies as graphs are available in our repository: \github.}\label{fn:topo} does not exactly match a conventional 2D mesh network, so the XY-routing algorithm \cite{xy-routing} cannot be applied.

\paragraph{Optimization ``Ours Aligned''}
In our first optimization, we retain radix-$4$ compute reticles but rotate the interconnect reticles by $90$ degrees and align them so that each interconnect reticle connects to up to six compute reticles (see \Cref{fig:system-visualizations}b).
This reduces both the average path length and the number of interconnect reticles, accelerating the manufacturing process.
The overlapping area available per vertical connector decreases from $214.5$~mm$^2$ to $45.5$~mm$^2$, but even with a conservative $10\,\mu m$ hybrid bond pitch, only $3.2$~mm$^2$ is needed to implement a bidirectional $2$~TB/s link at $1$~GHz, so the overlap is more than sufficient.
Each interconnect reticle provides eight vertical connectors and uses a fully connected intra-reticle topology of four routers with concentration $2$.

\paragraph{Optimization ``Ours Interleaved''}
This optimization slightly modifies the previous reticle placement by interleaving the interconnect reticles instead of aligning them (see \Cref{fig:system-visualizations}c), resulting in a distinct network topology\footnotemark[\value{footnote}].

\paragraph{Optimization ``Ours Rotated''}
We maximize the network radix of both compute and interconnect reticles.
By reducing the interconnect reticle size to $22.98 \times 32.53$~mm and rotating them by $45$ degrees, each interconnect reticle overlaps with up to seven compute reticles (see \Cref{fig:system-visualizations}d).
Although the overlapping area is smaller, it supports up to $6$~TB/s links (assuming a \gls{hb} pitch of $10\,\mu$m) with more than $10$~mm$^2$ available per vertical connector.
Unlike the previous placements with radix-$4$ compute reticles, this configuration increases the compute reticle radix to $7$, slightly enlarging its area (see \Cref{sec:evaluation-area} for area evaluation details).
Each interconnect reticle features seven vertical connectors and employs a fully connected topology of four routers with concentration $1$ or $2$.
While a formal optimality proof is beyond the scope of this paper, an exhaustive search over all integer reticle positions and rotations found no configuration with a higher radix than seven.

\vspace{-0.35em}
\subsection{Placements for \lol~Logic-on-Logic}
\label{ssec:opt-loi}

\paragraph{Constraints}
We again assume that all reticles on a given wafer are identical.
While for \loi~\gls{loi} systems, only the compute reticles needed to tessellate the wafer plane and the interconnect reticles could be placed with spaces in between, in \lol~\gls{lol} systems, both wafers contain compute reticles and must tessellate the plane to maximize integration density.

\paragraph{Baseline}
We use the same baseline placement as in \loi~\gls{loi} systems (see \Cref{fig:system-visualizations}a).
The only difference is that both wafers now contain identical radix-$4$ compute reticles.

\paragraph{Optimization ``Ours Contoured''}
Because \lol~\gls{lol} systems prohibit spacing between reticles to maximize integration density, the three \loi~\gls{loi} optimizations relying on gaps between interconnect reticles are not applicable.
We therefore propose a new radix-$5$ placement using contoured reticles on both wafers.
The lower wafer features $H$-shaped reticles, while the upper wafer uses plus-shaped reticles (see \Cref{fig:system-visualizations}e).
By aligning the centers of these shapes, each reticle connects to up to five reticles on the opposite wafer.
The placement illustrated in \Cref{fig:system-visualizations}e is schematic, with exaggerated contouring that makes the total reticle area appear much smaller than the reticle limit.
In practice, contouring is limited to the minimum required to achieve the target link bandwidth (e.g., for $2$~TB/s links, the reticle area equals $98.5\%$ of the reticle limit).

\section{Evaluation}
\label{sec:evaluation}

\subsection{Experiment Setup}
\label{sec:evaluation-methodology}

We use the cycle-accurate BookSim2 \cite{booksim2} \gls{noc} simulator to perform flit-level simulations of each wafer-scale architecture, providing zero-load latency, saturation throughput, and the average number of router-to-router hops per packet.
BookSim2 models wormhole routing with virtual-channel flow control and a four-stage router pipeline (routing, virtual-channel allocation, switch allocation, and crossbar traversal).
All simulations are repeated three times with different random seeds, and the results are averaged.
Area and power estimates are obtained using the Orion3.0 \cite{orion3} \gls{noc} power and area model.
Because Orion3.0 supports only up to 45~nm technology, we scale the area and power results to 7~nm using DeepScaleTool \cite{deepscaletool}.

\subsubsection{Architectural Parameters}
\label{sec:evaluation-parameters}

We model interconnects with bidirectional links providing $2$~TB/s bandwidth per direction at $1$~GHz, matching the link bandwidth in Tesla's Dojo \cite{tesla-dojo}.
Links are implemented as pipelined interconnects with one pipeline stage (register buffer) every $2$~mm of physical wire length.
Routers have a latency of four cycles.
Through extensive performance exploration across different input buffer sizes, we found that large wafer-scale architectures with long pipelined links require $32$ flit buffers to exploit the full throughput potential.
Given this large buffer requirement, we consider virtual channels too costly and therefore assume a single virtual channel per physical channel in all experiments.
However, our proposed network optimizations are fully compatible with configurations using multiple virtual channels.

\subsubsection{Workloads}
\label{sec:evaluation-workloads}

We use four synthetic traffic patterns: uniform (modeling all-to-all workloads such as \gls{moe} training \cite{alltoall-moe}), random permutation (modeling shuffle-style workloads such as FFT or sorting \cite{dally-book}), neighbor (modeling stencil workloads such as fluid dynamics simulations \cite{dally-book}), and tornado (modeling long-stride communication).

In addition, we leverage the ATLAHS \cite{atlahs} toolchain to collect GOAL \cite{goal} formatted traces from Llama-7B \cite{llama} training, and extend BookSim2 to replay these traces on our wafer-scale architectures.
These traces capture inter-GPU messages, message sizes, computation phase durations, and all dependencies between communication and computation events.
We collect traces for training on 20, 24, 40, 48, 52, 64, 96, and 124 GPUs to obtain a suitable trace for each architecture under evaluation, noting that a few GPUs may be idle in some configurations.
Messages, which can reach 1.8 MB, are split into 2 KB packets for network transmission.
Because BookSim2’s cycle-accurate flit-level model makes the simulation of a full \gls{llm} training epoch impractically slow, we instead run three independent simulations of eight hours each for every one of the 48 architecture–placement combinations, resulting in an average of 1.17 million packets transmitted per simulation.

\subsubsection{Metrics and Measurement Methodology (Synthetic Traffic)}
\label{sec:evaluation-metrics}

\paragraph{Latency}
We use a BookSim2 simulation at a low injection rate to measure the zero-load latency, configuring each link’s latency according to its physical length and assigning a latency of one cycle to each vertical connector between wafers.

\begin{figure}[h]
\vspace{-0.75em}
\centering
\captionsetup{justification=centering}
\includegraphics[width=0.95\columnwidth]{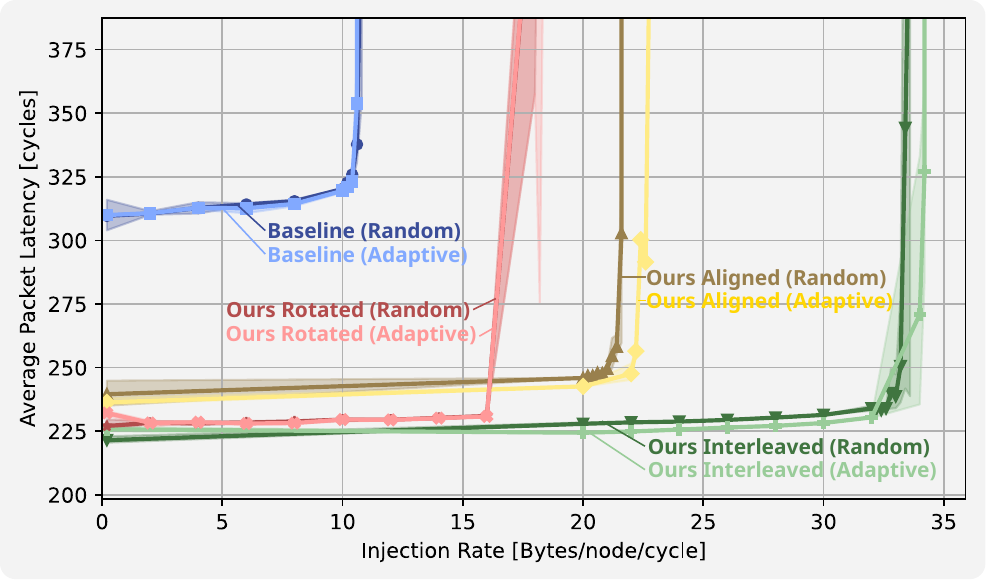}
\vspace{-1.0em}
\caption{{(\textsection \ref{sec:evaluation-latency-throughput}) Latency vs. Load for \loi~\gls{loi} with \wdl~300 mm wafers and \wum~maximized utilization (permutation traffic).}}
\label{fig:latency-vs-load}
\vspace{-2.0em}
\end{figure}

\paragraph{Throughput}
We progressively increase the injection rate in BookSim2 by $10\%$, $1\%$, $0.1\%$, and $0.01\%$ increments to accurately determine the network's saturation throughput (the point where the latency exceeds twice the zero-load latency).
\Cref{fig:latency-vs-load} shows a latency vs. load curve, where each point represents a BookSim2 simulation at a specific injection rate.

\paragraph{Area}
We use Orion3.0 to estimate the area of \gls{noc} routers, assuming that input buffers are implemented as SRAM rather than flip-flops.
Since SRAM scaling has plateaued compared to logic scaling, we apply a scaling factor of $0.2$ to scale SRAM area from $45$~nm to $7$~nm, which is more conservative than DeepScaleTool’s area scaling factor of $0.0271$.

\paragraph{Power and Energy}
We use Orion3.0 to estimate the power consumption of each \gls{noc} router.
For buffered links, we use BookSim2 results to estimate the average number of pipeline stages traversed per flit by subtracting the product of the average hop count and router latency from the zero-load latency.
We assume a conservative energy of $2$~pJ per bit per pipeline stage, consistent with prior work \cite{params-meta}.
Because the energy consumption of \glspl{hb} is negligible \cite{params-meta}, we do not model it explicitly.
Our analysis shows that in wafer-scale architectures, link power consumption exceeds router power by orders of magnitude, so we report only the total network power consumption without separating router and link power.

\subsection{Experiment Results on Synthetic Traffic}
\label{sec:evaluation-results-synthetic}

\subsubsection{Latency and Throughput}
\label{sec:evaluation-latency-throughput}

\begin{figure}[t]
\centering
\captionsetup{justification=centering}
\includegraphics[width=1.0\columnwidth]{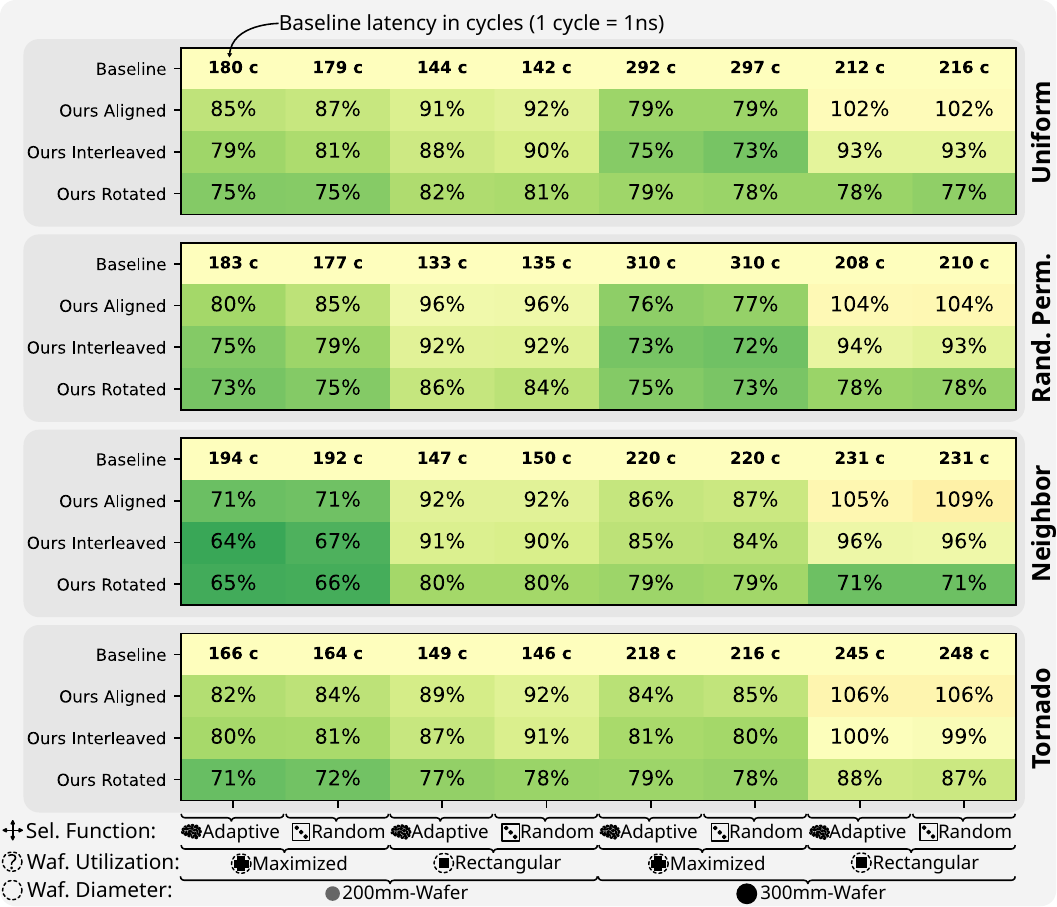}
\vspace{-2.25em}
\caption{{(\textsection \ref{sec:evaluation-latency-throughput}) Latency of \loi~Logic-on-Interconnect.}}
\label{fig:hm-latency-loi}
\vspace{-0.5em}
\end{figure}

\begin{figure}[t]
\centering
\captionsetup{justification=centering}
\includegraphics[width=1.0\columnwidth]{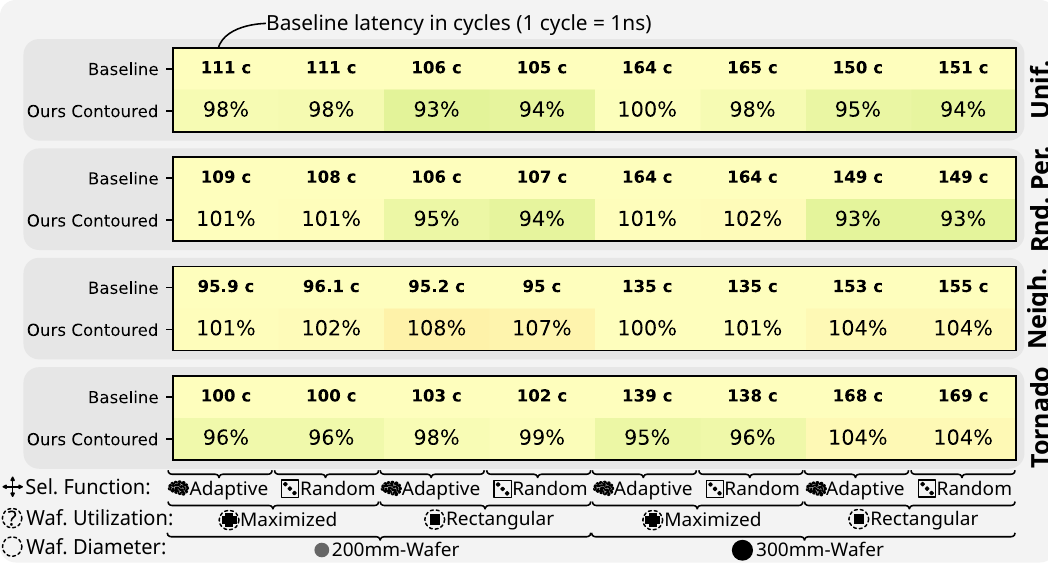}
\vspace{-2.25em}
\caption{{(\textsection \ref{sec:evaluation-latency-throughput}) Latency of \lol~Logic-on-Logic.}}
\label{fig:hm-latency-lol}
\vspace{-1.25em}
\end{figure}

\begin{figure}[t]
\centering
\captionsetup{justification=centering}
\includegraphics[width=1.0\columnwidth]{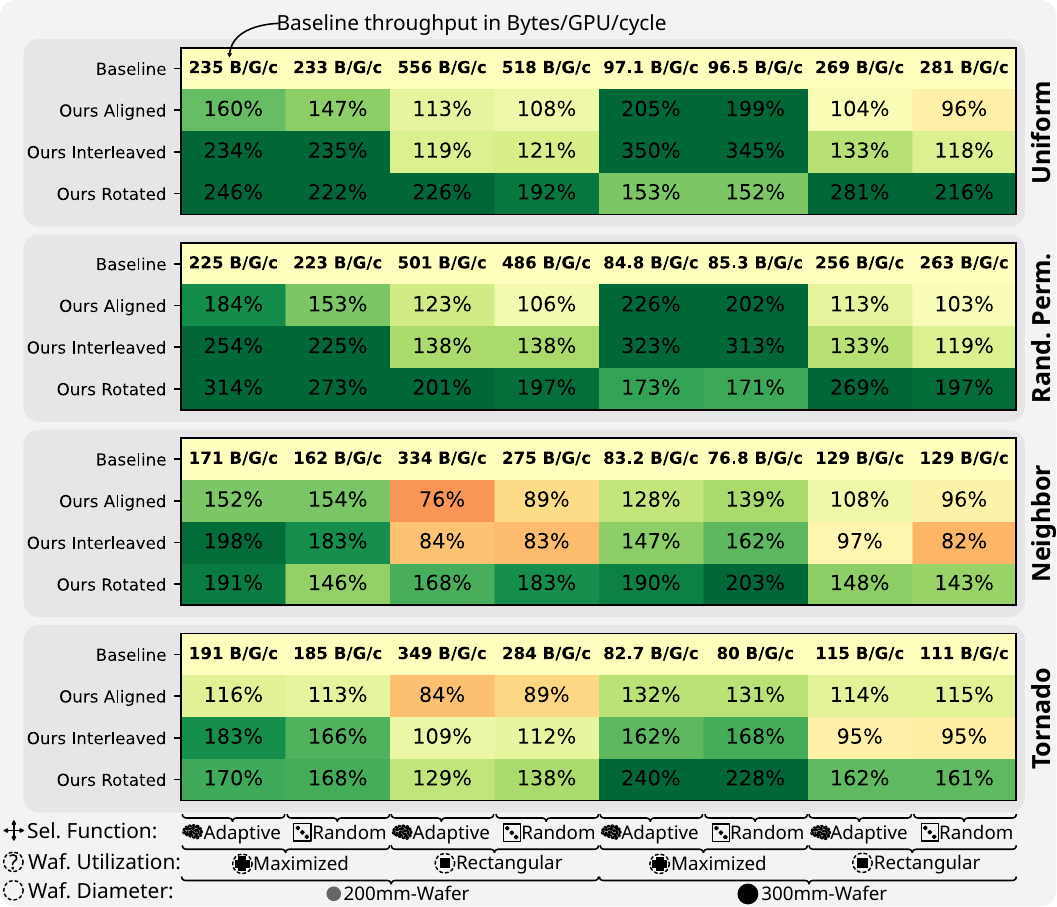}
\vspace{-2.25em}
\caption{{(\textsection \ref{sec:evaluation-latency-throughput}) Throughput of \loi~Logic-on-Interconnect.}}
\label{fig:hm-throughput-loi}
\vspace{-0.5em}
\end{figure}

\begin{figure}[t]
\centering
\captionsetup{justification=centering}
\includegraphics[width=1.0\columnwidth]{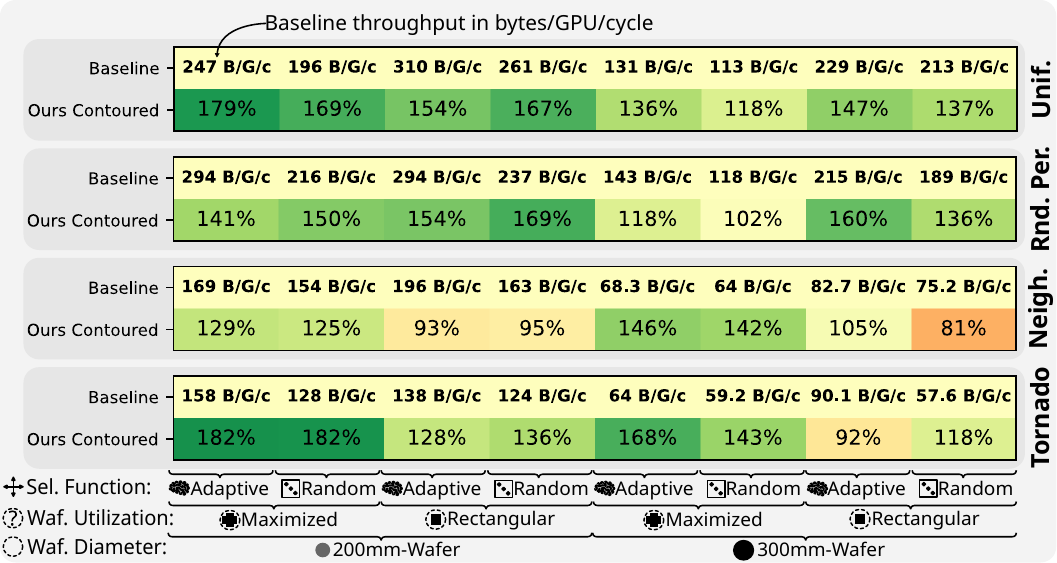}
\vspace{-2.25em}
\caption{{(\textsection \ref{sec:evaluation-latency-throughput}) Throughput of \lol~Logic-on-Logic.}}
\label{fig:hm-throughput-lol}
\vspace{-1.25em}
\end{figure}

\Cref{fig:latency-vs-load} shows detailed latency vs. load curves for our four reticle placements and two selection functions on the \loi~\gls{loi} system with \wdl~300~mm wafers and \wum~maximized wafer utilization under random permutation traffic.
Similar plots for the remaining $31$ experiments are available in our open-source repository\footnote{\github}.
Due to space constraints, we summarize the latency and throughput results in heatmap plots showing improvements over the baseline placement in \Cref{fig:hm-latency-loi,fig:hm-throughput-loi,fig:hm-latency-lol,fig:hm-throughput-lol}.
Analyzing these heatmaps provides insights into the performance of our proposed placements across system configurations and traffic patterns.

Our \textit{Aligned} and \textit{Interleaved} placements increase throughput while reducing latency across all systems with \wum~maximized wafer utilization.  
With \wur~rectangular utilization, they still improve these two metrics in most cases, though they may underperform the \textit{Baseline} for tornado and neighbor traffic.  
The \textit{Rotated} placement consistently outperforms the \textit{Baseline} across all architectures and traffic patterns.  
The \textit{Contoured} placement for \lol~\gls{lol} systems improves throughput in most cases while maintaining similar latency as the \textit{Baseline}.  
The \sfa~adaptive selection function slightly increases throughput at comparable latency to the \sfr~random selection function.  
Overall, our optimized placements achieve stronger improvements for \wum~maximized than for \wur~rectangular wafer utilization and for \loi~\gls{loi} than for \lol~\gls{lol} systems, with wafer diameter \wdi~having only a minor effect.  
Performance gains are also more consistent for random uniform and random permutation traffic than for tornado and neighbor traffic.

\subsubsection{Area}
\label{sec:evaluation-area}

\begin{figure}[b]
\vspace{-1.0em}
\centering
\captionsetup{justification=centering}
\includegraphics[width=1.0\columnwidth]{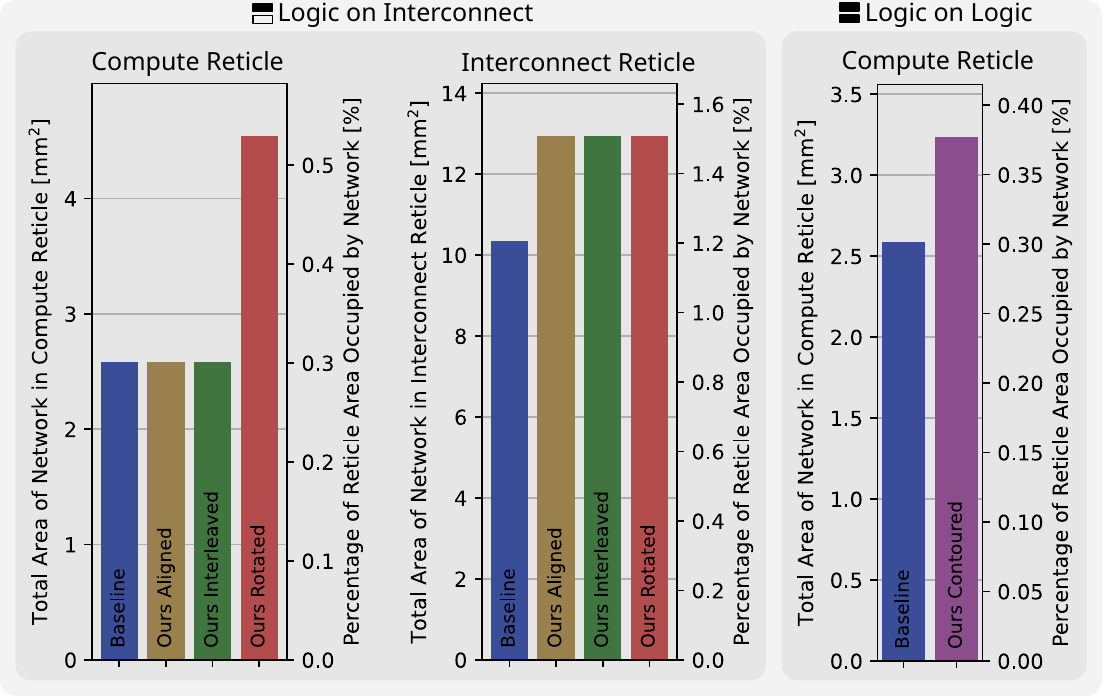}
\vspace{-2.25em}
\caption{{(\textsection \ref{sec:evaluation-area}) Area occupied by Network Routers.}}
\label{fig:area-analysis}
\end{figure}

\Cref{fig:area-analysis} shows the area occupied by routers on compute and, where applicable, interconnect reticles.
Because we report per-reticle area, these results are independent of \wdi~wafer diameter and \wu~utilization and therefore apply to all \loi~\gls{loi} and \lol~\gls{lol} systems.
We observe that routers occupy only a small fraction of the reticle area, and our proposed placements introduce little or no additional area overhead compared to the baseline.
Router area is dominated by input buffers, with the remaining router logic contributing negligibly.
A detailed study of reticle wiring resources is beyond the scope of this work, but global wiring usage can be expected to scale roughly with router area.

\subsubsection{Power and Energy}
\label{sec:evaluation-power-energy}

\begin{figure}[h]
\centering
\captionsetup{justification=centering}
\includegraphics[width=1.0\columnwidth]{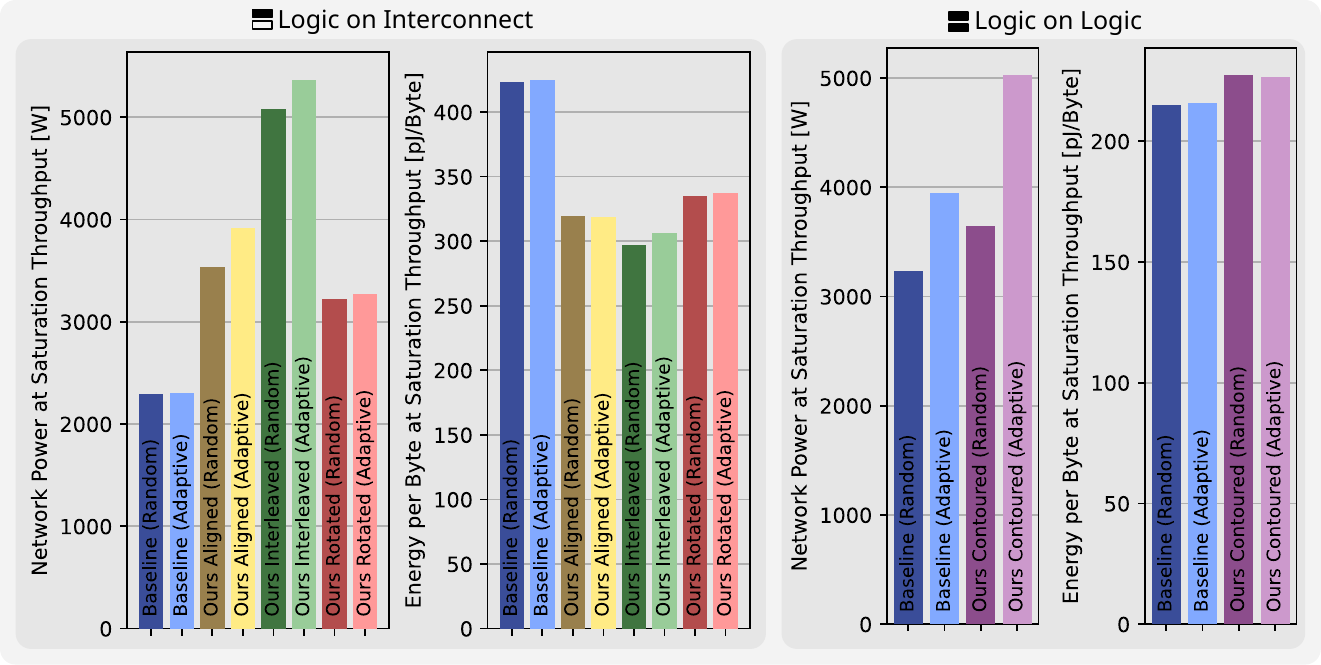}
\vspace{-2.25em}
\caption{{(\textsection \ref{sec:evaluation-power-energy}) Power \& energy for \loi~\gls{loi} with \wdl~300 mm wafers and \wum~maximized utilization (permutation traffic).}}
\label{fig:power-analysis}
\vspace{-1.25em}
\end{figure}

\begin{figure}[t]
\centering
\captionsetup{justification=centering}
\includegraphics[width=1.0\columnwidth]{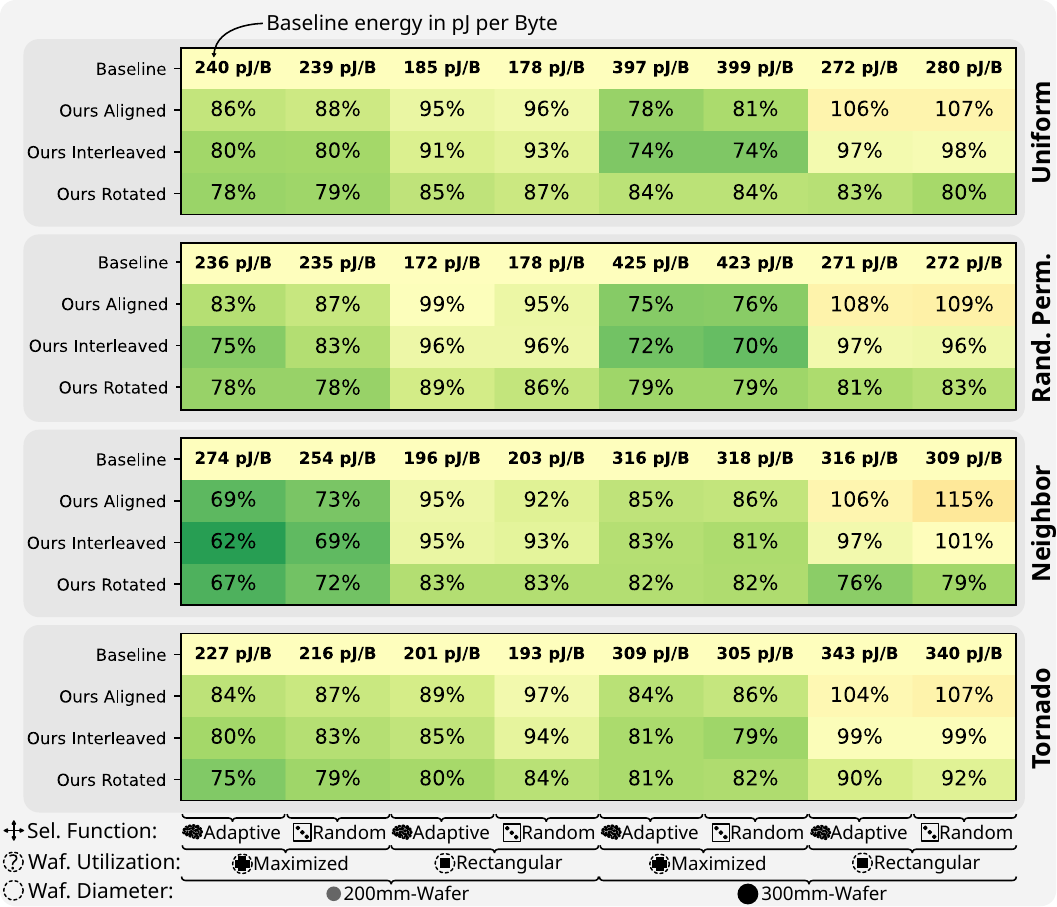}
\vspace{-2.25em}
\caption{{(\textsection \ref{sec:evaluation-power-energy}) Energy of \loi~Logic-on-Interconnect.}}
\label{fig:hm-energy-loi}
\vspace{-1.25em}
\end{figure}

\begin{figure}[!h]
\centering
\captionsetup{justification=centering}
\includegraphics[width=1.0\columnwidth]{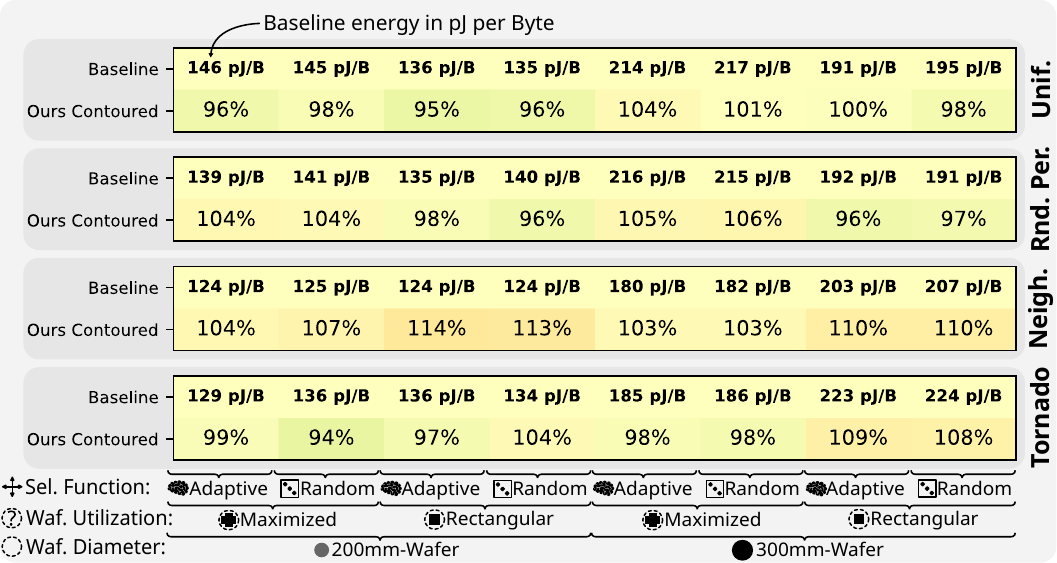}
\vspace{-2.25em}
\caption{{(\textsection \ref{sec:evaluation-power-energy}) Energy of \lol~Logic-on-Logic.}}
\label{fig:hm-energy-lol}
\vspace{-1.25em}
\end{figure}

\Cref{fig:power-analysis} shows the total power consumption of the wafer-scale network (left) and the normalized energy per transferred byte (right) at saturation throughput for the two \sfi~selection functions on the \loi~\gls{loi} system with \wdl~300~mm wafers and \wum~maximized wafer utilization under random permutation traffic.
Equivalent plots for the remaining $31$ experiments are available in our open-source repository\footnote{\github}.
We summarize the energy efficiency results in \Cref{fig:hm-energy-loi,fig:hm-energy-lol}.
Comparing the network power of about 4 kW to the reported 15 kW wafer-scale power budget \cite{fred} suggests that up to one quarter of the total power could be devoted to global data movement.
Note that power is evaluated at saturation throughput.
Since average network utilization is typically well below 100\%, actual power consumption under normal operation will be much lower.
The energy per byte shows that our optimized placements typically improve efficiency by shortening the average path length, while the higher total power mainly results from increased saturation throughput.

\subsection{Experiment Results on Application Traces}
\label{sec:evaluation-results-traces}

\begin{figure}[h]
\vspace{-1.0em}
\centering
\captionsetup{justification=centering}
\includegraphics[width=1.0\columnwidth]{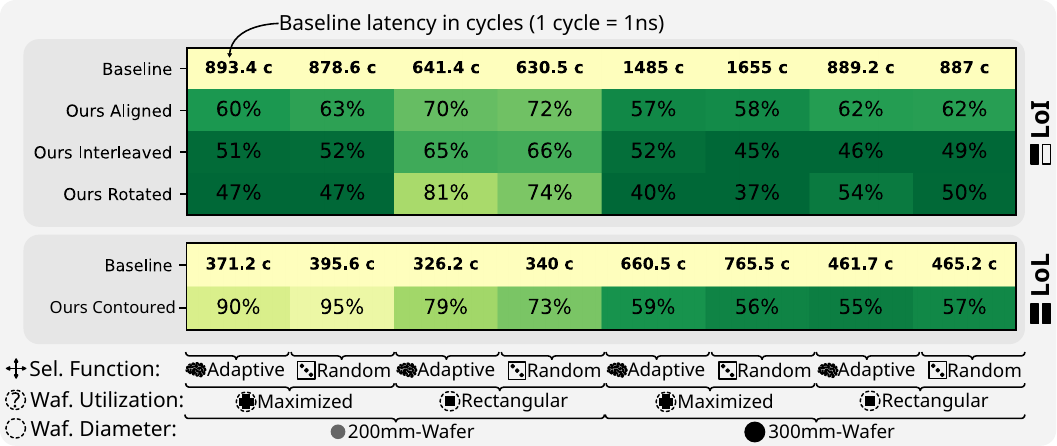}
\vspace{-2.25em}
\caption{{(\textsection \ref{sec:evaluation-results-traces}) Network latency during trace simulation.}}
\label{fig:trace-latency}
\vspace{-1.5em}
\end{figure}

\Cref{fig:trace-latency} shows the average network latency observed when running the Llama-7B training traces.
During these simulations the network alternates between phases of high load with severe congestion and phases of lower load where computation dominates.
This elevated congestion leads to substantially higher average latencies than in the synthetic traffic experiments where the zero-load latency captures the latency without contention.
Our results indicate that for \gls{llm} training workloads the latency reductions achieved by our proposed placements exceed those measured with synthetic traffic.
On average latency decreases to 60\% of the baseline and in the best case to 37\%.
While beneficial for all architectures, our optimized placements yield larger improvements for \gls{llm} training on \wdl~300~mm wafers than on \wds~200~mm wafers, and on \loi~\gls{loi} systems than on \lol~\gls{lol} systems.

\section{Related Work}
\label{sec:rl}

With its SoIC technology, TSMC offers wafer-on-wafer hybrid bonding with fine-pitch connections, and its roadmap \cite{tsmc-soic-wow} projects a $2\times$ increase in interconnect density every two years.
Such rapid scaling has become feasible only through recent advances in hybrid bonding processes \cite{stacking-3-wafers}, comprehensively reviewed by Lau et al. \cite{hybrid-bonding}.
For a broader overview of wafer-scale computing, we refer to Hu et al. \cite{wafer-scale-survey}.

While no prior work has addressed network design for \gls{wsi} systems based on wafer-on-wafer hybrid bonding, several studies have explored related directions for chiplet-based \gls{wsi}.
FRED \cite{fred} employs a Clos-like topology for wafer-scale systems to accelerate collective operations in DNN training, but such topologies are infeasible under the geometric constraints of wafer-on-wafer hybrid bonding.
Network-on-Wafer \cite{network-on-wafer} co-designs topology, routing, and collective operations for wafer-scale systems and assumes per-link bandwidths of $2$~TB/s, similar to Tesla Dojo \cite{tesla-dojo} and our work.
WSC-LLM \cite{wsc-llm} explores joint architectural and scheduling optimization using a 2D mesh topology for inter-die communication.
Other studies adopting 2D mesh topologies and motivating our mesh-like baseline include the wafer-scale AI accelerator Simba \cite{simba}, a wafer-scale GPU architecture \cite{waferscale-gpu}, and a 2048-chiplet wafer-scale system developed by UCLA and UIUC \cite{ucla-uiuc}.

\section{Conclusion}
\label{sec:conc}

Wafer-on-wafer hybrid bonding is a promising and readily available technology for realizing \gls{wsi} with high-bandwidth interconnects.
The constraint that network topology must emerge from connecting overlapping reticles on opposite wafers creates a new and unexplored design space.
In this work, we explore said design space by optimizing the placement of reticles on the top and bottom wafers to maximize the number of neighbors per reticle and thereby minimize the network’s average path length.

Our comprehensive evaluation shows that our proposed reticle placements significantly improve throughput, latency, and energy efficiency across almost all integration levels, wafer diameters, wafer utilizations, and workloads considered.  
We achieve throughput improvements of up to $250\%$, latency reductions of up to $36\%$, and energy reductions of up to $38\%$.

\ifnb
\section*{Acknowledgements}
\label{sec:ack}

We thank Shuhao Li, Timo Schneider, and Siyuan Shen for their support with the ATLAHS toolchain and we thank Jiho Kim for insightful discussions on area and power modeling.
Furthermore, we thank the Swiss National Supercomputing Centre (CSCS) for access to their Alps system.
This work was supported by the ETH Future Computing Laboratory (EFCL), financed by a donation from Huawei Technologies.
It also received funding from the European Research Council \raisebox{-0.25em}{\includegraphics[height=1em]{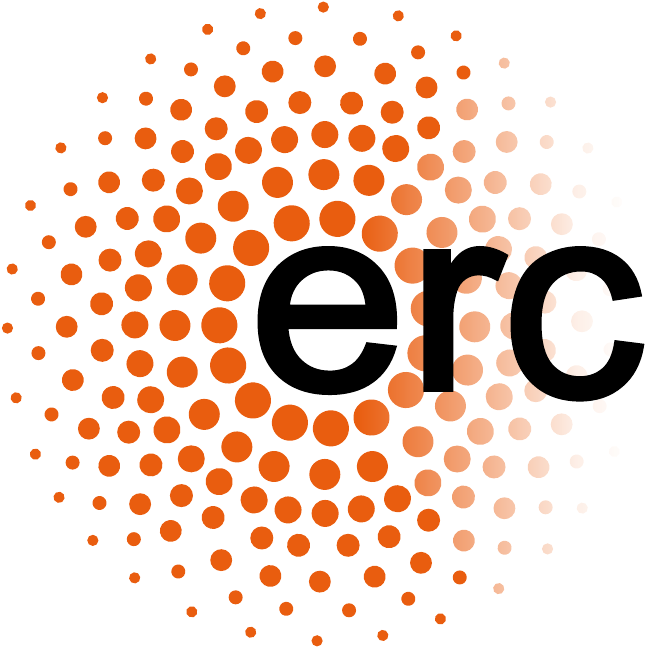}} (Project PSAP, No.~101002047) and from the European Union's HE research and innovation programme (grant agreement No.~101070141; Project GLACIATION).

\fi

\balance
\bibliographystyle{ACM-Reference-Format}
\bibliography{bibliography}

\end{document}